\documentstyle[sprocl]{article}
\def\be{\begin{equation}}
\def\ee{\end{equation}}

\begin{document}

\title{CHARGE DISTRIBUTION IN CHIRAL PION BREMSSTRAHLUNG}

\author{IGOR ANDREEV}

\address{ Lebedev Physical Institute,Leninsky pr.53,117924 Moscow,Russia}

\maketitle\abstracts{
It is argued that distribution over the number of charged and neutral
soft chiral pions are very broad if they are emitted coherently.}

\section{Introduction}

Charge distribution of pions in multiple production processes
drew much attention recently. The growth of interest in this
subject is due to expectations to detect the disoriented chiral
condensate (DCC) formation in high energy collisions~\cite{1}$^{-}$\cite{6}.
The simplest picture of the process is given by "Baked Alaska
scenario"~\cite{4}, where coherent pulses of semiclassical pion
field are emitted leading to anomalously large
fluctuations in the ratio of neutral to charge pions produced.
In particular, the probability to produce $n_o$ neutral pions
(for large total number of pions $n$) is given by inverse square
root formula,
\begin{equation}
w(n_0)\sim 1/\sqrt{n_0 n} ,
\label{1}
\end{equation}
being very flat and so quite different from usual binomial-like
distributions. This mechanism may be relevant for description of
"Centauro" (and "Anti-Centauro") type events found in cosmic ray
experiments, see~\cite{7,8} and references therein, in which the number
of charged particles drastically exceeds the number of neutral ones
(or vice versa).

\section{Coherent pion production with isospin conservation}

Now the problem arises -- to what extent the behaviour in Eq.~(\ref{1})
can be considered as a signature of DCC formation. Let us remind in
this connection that the distribution of the form of Eq.~(\ref{1})
was found long ago in a model of independent pion production when
isotopic spin conservation was taken into account~\cite{9,10}.The role of
isotopic spin effects was analyzed within the framework of the model for
production of the coherent state of pions.If the different kinds of pions were
not coupled to each other,then we would have the following expression for the
final state:
  \be
 \vert f>=e^{-c/2}\exp[\int d^{3}k\sum_{i=1}^3
f_i(k)a_i^{\dag}(k)]\vert0>, \nonumber\\ \quad
c=\sum_i c_i=\sum_i\int d^3 k\vert
f_i(k)\vert^2
 \label{eq:2}
 \ee
 where $a_i^{\dag}$ are the creation operators
for $i$-type pions,$\vert0>$ is the pion vacuum and $\vert f(k)\vert^2$
determines the momentum distribution of pions [$f_i(k)$ is the effective
density of the pion source].The probability for production of $\pi^{\pm}$ and
$\pi^0$ mesons in this case would have the same form as in the model of
uncorrelated particle production and a Poisson distribution with respect to the
number of pions of each type would exist.  Let us now take into account that
$f_i$ and $a_i$ are components of vectors in the isotopic space.The state
$\vert f>$ in Eq.~\ref{eq:2} does not have a specific isospin $I$ (or electric
charge),but contains states with large values of $I$ which increase with
increasing number of particles.This is not consistent with the conservation of
total isospin in the collision process.For example, the isospin of a pion
system cannot be higher than 2 in the collision of two nucleons with production
of pions.  The state with isospin $I=0$ appears as a result of averaging $\vert
f>$ over directions of the ${\bf f}$ vector in the isospace,
 \be
 \vert f;I=0>=e^{-c/2}\int d{\Omega}\exp[\int d^{3}k{\bf f}(k){\bf a}^{\dag}
(k)]\vert0>
\label{3}
\ee
This state may be considered as a result of generating of isoscalar pairs
of pions:
\be
\vert f;I=0>=\frac{e^{-c/2}}{\sqrt{4\pi}}\sum_{m=0}^{\infty}\frac{1}
{(2m+1)!}(\sum_{j=1}^{3}[\int d^{3}kf(k)a_j^{\dag}(k)]^2 )^{m}\vert0>
\label{4}
\ee
It is the proper state for annihilation operator of isoscalar pair
and it may be considered as a coherent state of isoscalar pairs.

The important point is that the distributions with respect to
the number of neutral pions $w_{0}(n_{0})$ and charged pions $w_{ch}(n_{ch})$
for $I=0$ are much broader than those obtained by ignoring the constraints
imposed on the isospin of the pion system,if the total average number
of pions is large,$<n>\gg1$.In the main region of variation of $n_0$
we obtain
\be
w_{0}(n_{0},I=0)\cong\frac{1}{\sqrt{2c}}\frac{\Gamma(\frac{n_{0}+1}{2})}
{\Gamma(\frac{n_0}{2}+1)},\quad <n_{0}>=c/3
\label{5}
\ee
i.e.,for example,the probability for occurence of events without $\pi^{0}$
is high enough,amounting here to more than 10 per cent for $<n>\approx c=100$.
The probability $w_{0}(n_{0},I=0)$ decreases slowly,in accordance with
Eq.~\ref{1},up to $n_{0}\sim c$ with increasing $n_{0}$,where
$w_{0}(c,I=0)=1/2c$ and then decreases to zero in the region
$\vert n_{0}-c\vert\sim\sqrt c$.
Such behaviour of $w_{0}$ can explain an essential fraction of Centauro-type
events in which the $\pi^0$ mesons are missing at $<n>\gg 1$.The probability
for the occurence of such events would be $e^{-c/3}\sim e^{-33}<10^{-14}$
for the original Poisson distribution,i.e.,they would not be observed.

Analogously we can see that for $c-n_{ch}\gg\sqrt c$ the distribution of
charged pions is given by
\be
w_{ch}(n_{ch},I=0)\cong1/\sqrt{c(c-n_{ch})},\quad <n_{ch}>\cong2c/3,
\label{6}
\ee
we again obtain a slowly varying function of $n_{ch}$.The probaility
$w_{ch}(n_{ch},I=0)$ increases slowly with increasing $n_{ch}$,reaching a
maximum at $n_{ch}\cong c$
\be
w_{ch}(c,I=1)\cong\frac{\Gamma(1/4)}{\sqrt{2\pi}(2c)^{3/4}}
\label{7}
\ee
and then decreases fastly in the interval $\vert n_{ch}-c\vert\sim\sqrt c$.
The events without charged pions also comprise a sizable fraction,
$w_{ch}(0,I=0)\cong1/c=0.01$.

The same qualitative results were obtained when the states with $I=1$ were selected.
More recently similar results were obtained for squeezed states~\cite{11}.

\section{Soft chiral pion bremsstrahlung}

Below we consider soft chiral pion bremsstrahlung
accompanying some basic high-energy
process and estimate charge distribution of the pions~\cite{12}. The
quantum charge states of chiral pions emitted from simple vertices will be
explicitly calculated. It will be found that neutral pion number distribution
again has the form of Eq.~(\ref{1}). That is, such flat charge
distributions are typical for soft chiral pions and do not indicate
directly on DCC formation.

The soft chiral pion bremsstrahlung was studied many years ago~\cite{13,14}.
Similarly to photons, soft pions are emitted from external lines of
diagrams representing the basic process (to be definite, we shall take
external particles to be spin 1/2 fermions (nucleons)). The complications
arising due to noncommutative pion-nucleon vertices and nonlinear pion-nucleon
coupling were shown to be mutually cancelled~\cite{13}. Nonlinear pion-pion
coupling can be taken into account but its effect vanishes in the limit of
small pion momenta. Soft virtual pion exchange~\cite{15} changes normalization
and does not influence the pion number distributions. The net result for
soft pion emission is given by substitution~\cite{14}:
\begin{equation}
\psi_j \rightarrow \exp(-i\gamma_5\tau_i\phi_i/2)\psi_j,
\label{8}
\end{equation}
where $\psi_j$ is the fermion field for every incoming or outgoing
nucleon in the skeleton diagram of a basic process, $\phi_i=\pi_i/f_\pi$,
$\pi_i$ being the pion field, $f_\pi=93$ MeV is the pion decay constant.

As the simplest example consider the scalar vertex $\bar{\psi}\psi$
($\Gamma$ is the identity matrix). Its chiral-invariant extension has the
form

\begin{equation}
V_s = \bar{\psi} \exp(-i\gamma_5\tau_i\phi_i)\psi(x).
\label{9}
\end{equation}

\noindent
It coincides formally with the modified nucleon mass term in the chiral
lagrangian. We neglect pion momenta and for the fields $\phi_i$ use
the decomposition

\begin{equation}
\phi_i\rightarrow \phi_{i}(0) = \phi_{i}^{+} + \phi_{i}^{-} =
\int d^3 k f(k)[a^{+}_{i}(k) + a^{-}_{i}(k)],
\label{10}
\end{equation}

\noindent
where creation and annihilation operators $a^{+}_{i}(k), a^{-}_{i}(k)$
obey canonical commutation relations. In the free field approximation

\begin{equation}
f(k)=(2\pi)^{-3/2}(2k_0)^{-1/2}f^{-1}_{\pi}, \quad k_0\le k_{m}
\label{11}
\end{equation}
where $k_{m}$ is an upper limit of pion softness.Then we must
 calculate the matrix elements of $\pi^{0}, \pi^{+}, \pi^{-}$
production

\begin{equation}
M_s=\langle n_{+},n_{-},n_{0}|\exp(-i\gamma_5\tau_i\phi_{i}(0))|0\rangle
\label{12}
\end{equation}
The total isotopic spin of the pions produced can be zero or one.
Consider the first case.
If the average number of pions is large (it is the most interesting
case) then

\begin{equation}
\langle n\rangle \cong c,\qquad c=\int d^{3}k|f(k)|^{2}\gg 1.
\label{13}
\end{equation}
To estimate it take the free field approximation (\ref{11});
then

\begin{equation}
\langle n\rangle = \frac{1}{f^2_{\pi}}\int\frac{d^3 k}{(2\pi)^3 2k_0}
\cong  \frac{k_{m}^2}{8\pi^2 f^2_{\pi}}
\label{14}
\end{equation}
A prominent feature of the model is the distribution over number
of charged and neutral pions produced. It can be obtained from
matrix elements (\ref{12}) and has the multiplicative form with
respect to the total number of pions $n=n_0+n_{ch}$ and the number
of neutral pions $n_0$,

\begin{equation}
w(n,n_0) = w(n)w_n(n_0)
\label{15}
\end{equation}
where

\begin{equation}
w_n(n_0) = \frac{1}{n+1} \frac{(n/2)!}{\Gamma(\frac{n+1}{2})}
\frac{\Gamma(\frac{n_0+1}{2})}{(n_0/2)!}
 \approx \frac{1}{\sqrt{n n_0}}
\label{16}
\end{equation}
is the probability to produce $n_0$ neutral pions for the given
total number of pions ($\Gamma(n)$ is the Euler $\Gamma$-function,
$n_0$ and $n$ are even, $n_0\le n$), and

\begin{equation}
w(n) = \frac{(n-c+1)^2}{N_s}
\frac{e^{-c}c^n}{(n+1)!}
\label{17}
\end{equation}
is the distribution over the total number of pions produced.

The distribution (\ref{16}) over the number of neutral pions $n_0$
and corresponding distribution over the number of charged pions
are very broad.  These distributions appear to be very similar to those
given by Eqs.\ref{5}-\ref{6}.The distribution
(\ref{17}) over the total number of pions is Poisson-like (though with
an additional central dip at $n\sim\langle n\rangle$) and it is much
more narrow than (\ref{16}).

As a case of immediate physical interest consider now the soft pion emission
for electromagnetic scattering of strongly interaction fermions.
The skeleton vertex has now the form

\begin{equation}
V_0 = e \bar{\psi}\gamma_{\mu}Q\psi =
e \bar{\psi}\gamma_{\mu}\frac{\tau_3+N_B}{2}\psi
\label{18}
\end{equation}
where the baryon number $N_B = 1$ for nucleons and $Q$ is electric
charge in units of $e$. Chiral extension of the vertex is taken as

\begin{equation}
V = e \bar{\psi}\exp(-i\gamma_5\tau_k\phi_k/2)\gamma_{\mu}
\frac{\tau_3+N_B}{2}\exp(-i\gamma_5\tau_k\phi_k/2)\psi
\label{19}
\end{equation}
We consider diagonal transitions with high average multiplicity,$c\gg1$.
 Then the average number of pions is

\begin{equation}
<n>=\frac{1}{3}c,\quad c\gg1
\label{20}
\end{equation}
and the distribution over the number of neutral and charged pions is given by

\begin{eqnarray}
w_2(n,n_0)  \cong
\frac{3(c-n_0)^2}{2N_ec^{2}\sqrt{2c}}
\frac{\Gamma(\frac{n_0+1}{2})}{\left(\frac{n_0}{2}\right)!}
        w_2(n)
\label{21}
\\
w_2(n)  \cong
\frac{2}{3c\sqrt{2\pi c}}(n-c)^2
\exp\left(-\frac{(n-c)^2}{2c}\right)
\label{22}
\end{eqnarray}
where $w_2(n)$ is the probability to find $n=n_0+n_c$ pions, $n$
and $n_0\le n$ are even.

The distribution over the number of neutral (or charged) pions in
Eq.~(\ref{21}) is again very broad ensuring a sizable number of
events, in which almost all pions are neutral (or charged).
The distribution $w_2(n)$ over the total number of pions is
again narrow and in fact coincides with Eq.~(\ref{17}) for
$c\gg 1$, $n\gg1$ up to a normalization factor. The total probability
of high multiplicity events for electromagnetic vertex is

\begin{equation}
\sum_{n=2k} w_2(n)=\frac{1}{3}
\label{23}
\end{equation}
just corresponding to factor $1/3$ in Eq.~(\ref{20}).

\section{Discussion}

Examples of coherent soft pion emission considered above show very broad
distributions over the number of neutral and charged pions in high
multiplicity events.The neutral pion distributions in all cases are
given essentially by the function of Eq.~\ref{5} leading to inverse
square root behaviour of Eq.~\ref{1} characteristic for DCC formation.
So we conclude that one has to look for more delicate criteria of DCC
formation.Here the momentum distribution of pions produced may appear
helpful.

The bremsstrahlung spectrum of pseudoscalar pions has the form
$dn\sim kdk$ (contrary to photon spectrum $dk/k$) and so very small
momenta $k$ are inefficient for this mechanism.It was necessary
(as in Bloch-Nordsiek model) to introduce an upper limit of pion softness,
$k<k_{m}$ and the total number of pions produced by this mechanism
is proportional to $k_{m}^2$.The value of $k_m$ is not quite definite
(the most severe possible estimation is around rho-meson mass)
but it does not exceed the momentum transfer $\Delta p$ in the baryonic vertex
$\Gamma$.Anyhow it is clear that the presence of large baryonic momentum
transfer $\Delta p$ (and so the presence of high $p_{T}$ baryons)
is highly favourable for copious production of pions by the bremsstrahlung
mechanism.At the same time the soft pions are expected to be present
in lower $p_{T}$ region.In the last region the pion spectrum of the form
$kdk$ by itself can be used for identification of the process of pion
bremsstrahlung.It can be seen in future experiments when it will be possible
to look at narrow windows of $p_T$.

The conditions for such broad distributions to appear in high
multiplicity events are small isotopic spins of the pion system
and many particle matrix elements symmetric with respect to pion
momenta, thus ensuring a constructive interference.
This can be seen already from an early paper by A.~Pais~\cite{16}
and was explicitly demonstrated more recently in paper \cite{17}.
Both of these conditions are fulfilled in our model examples.

In conclusion, it thus appears that inverse square root
distributions over number of neutral and charged pions are of
very general nature being characteristic for coherent soft
pion radiation.

\section*{Acknowledgements}
This work was supported by Russian Fund for Fundamental Research
(grant 96-02-16210a).

\end{document}